\documentclass[a4paper,12pt]{article}
\usepackage{graphicx, rotating}

\ifx\pdfoutput\undefined
\usepackage[dvips,bookmarks]{hyperref}	
\else
\usepackage{hyperref}	
\fi
\hypersetup{colorlinks,bookmarksopen,bookmarksnumbered,citecolor=verdes,
linkcolor=blus,pdfstartview=FitH,urlcolor=rossos}
\def\hhref#1{\href{http://arxiv.org/abs/#1}{#1}} 

\font\mini=cmr10 at 2pt

\usepackage{multicol}
\usepackage{color}
\definecolor{rosso}{cmyk}{0,1,1,0.4}
\definecolor{rossos}{cmyk}{0,1,1,0.55}
\definecolor{rossoc}{cmyk}{0,1,1,0.2}
\definecolor{blu}{cmyk}{1,1,0,0.3}
\definecolor{blus}{cmyk}{1,1,0,0.6}
\definecolor{bluc}{cmyk}{1,1,0,0.1}
\definecolor{verde}{cmyk}{0.92,0,0.59,0.25}
\definecolor{verdec}{cmyk}{0.92,0,0.59,0.15}
\definecolor{verdes}{cmyk}{0.92,0,0.59,0.4}
\newcommand{\diag}{\hbox{diag}\,}
\font\tenrsfs=rsfs10 at 12pt
\font\sevenrsfs=rsfs7
\font\fiversfs=rsfs5
\newfam\rsfsfam
\textfont\rsfsfam=\tenrsfs
\scriptfont\rsfsfam=\sevenrsfs
\scriptscriptfont\rsfsfam=\fiversfs
\def\mathscr#1{{\fam\rsfsfam\relax#1}}
\def\Lag{\mathscr{L}}

\newcommand{\fig}[1]{~\ref{fig:#1}}
\newcommand{\riga}[1]{\noalign{\hbox{\parbox{\textwidth}{#1}}}\nonumber}

\newcommand{\eq}[1]{~{\rm (\ref{eq:#1})}}

\newcommand{\GeV}{\,{\rm GeV}}
\newcommand{\TeV}{\,{\rm TeV}}
\newcommand{\eV}{\,{\rm eV}}

\def\circa#1{\,\raise.3ex\hbox{$#1$\kern-.75em\lower1ex\hbox{$\sim$}}\,}
\newcommand{\nubarnu}{\raisebox{1ex}{\hbox{\tiny(}}\overline\nu\raisebox{1ex}{\hbox{\tiny)}}\hspace{-0.5ex}}

\newcommand{\DM}{{\rm DM}}

\newcommand{\ds}{\partial\hspace{-1.2ex}/\hspace{0.3ex}}

\newcommand{\NP}{Nucl. Phys.}

\newcommand{\PL}{Phys. Lett.}
\newcommand{\PR}{Phys. Rev.}

\newcommand{\beq}{\begin{equation}}
\newcommand{\eeq}{\end{equation}}
\newcommand{\MeV}{\,{\rm MeV}}
\def\circa#1{\,\raise.3ex\hbox{$#1$\kern-.75em\lower1ex\hbox{$\sim$}}\,}
\makeatletter

\def\art{\@ifnextchar[{\eart}{\oart}}
\def\eart[#1]#2#3#4#5#6{{\rm #2}, {#3 #4} {\rm (#6) #5} [{\hhref{#1}}]}
\def\hepart[#1]#2{{\rm #2, \hhref{#1}}}
\newcommand{\oart}[5]{{\rm #1}, {#2 #3} {\rm (#5) #4}}

\newcounter{alphaequation}[equation]
\def\thealphaequation{\theequation\hbox to
0.6em{\hfil\alph{alphaequation}\hfil}}
\def\eqnsystem#1{
\def\@eqnnum{{\rm (\thealphaequation)}}
\def\@@eqncr{\let\@tempa\relax \ifcase\@eqcnt \def\@tempa{& & &} \or
  \def\@tempa{& &}\or \def\@tempa{&}\fi\@tempa
  \if@eqnsw\@eqnnum\refstepcounter{alphaequation}\fi
\global\@eqnswtrue\global\@eqcnt=0\cr}
\refstepcounter{equation} \let\@currentlabel\theequation \def\@tempb{#1}
\ifx\@tempb\empty\else\label{#1}\fi
\refstepcounter{alphaequation}
\let\@currentlabel\thealphaequation
\global\@eqnswtrue\global\@eqcnt=0 \tabskip\@centering\let\\=\@eqncr
$$\halign to \displaywidth\bgroup \@eqnsel\hskip\@centering
$\displaystyle\tabskip\z@{##}$&\global\@eqcnt\@ne
\hskip2\arraycolsep\hfil${##}$\hfil& \global\@eqcnt\tw@\hskip2\arraycolsep
$\displaystyle\tabskip\z@{##}$\hfil
\tabskip\@centering&\llap{##}\tabskip\z@\cr}
\def\endeqnsystem{\@@eqncr\egroup$$\global\@ignoretrue} \makeatother

\newcommand{\SU}{\rm SU}

\begin{document}\begin{center}
{DFTT40/2005 \hfill IFUP--TH/2005-34}
\color{black}
\vspace{1.0cm}

{\Huge\bf\color{rossos}Minimal Dark Matter}
\medskip
\bigskip\color{black}\vspace{0.6cm}

{
{\large\bf Marco Cirelli}$^a$,
{\large\bf Nicolao Fornengo}$^b$,
{\large\bf Alessandro Strumia}$^c$.
}
\\[7mm]
{\it $^a$ Physics Department, Yale University, New Haven, CT 06520, USA}\\[3mm]
{\it $^b$ Dipartimento di Fisica Teorica, Universit\`a di Torino\\ and INFN, Sez.\ di Torino, via P. Giuria 1, I-10125 Torino, Italia}\\[3mm]
{\it $^e$ Dipartimento di Fisica dell'Universit{\`a} di Pisa and INFN, Italia}\\
\end{center}

\bigskip\bigskip\bigskip

\centerline{\large\bf\color{blus} Abstract}
\begin{quote}
\vspace{-0.25cm} \indent\color{blus}\large A few multiplets that can
be added to the SM contain a lightest neutral component which is
automatically stable and provides allowed DM candidates with a
non-standard phenomenology.  Thanks to coannihilations, a successful
thermal abundance is obtained for well defined DM masses.  The best
candidate seems to be a $\SU(2)_L$ fermion quintuplet with mass 4.4
TeV, accompanied by a charged partner $166\MeV$ heavier with life-time
$1.8\,{\rm cm}$, that manifests at colliders as charged tracks
disappearing in $\pi^\pm$ with $97.7\%$ branching ratio.
The cross section for usual NC direct DM detection is 
$\sigma_{\rm SI} = f^2\, 1.0\cdot 10^{-43}\,{\rm cm}^2$
where $f\sim 1$ is a nucleon matrix element.
We study
prospects for CC direct detection and for indirect detection.
\color{black}
\end{quote}

\bigskip

\section{Introduction}
The Dark Matter (DM) problem calls for new physics beyond the Standard Model (SM).
Its simplest interpretation consists in assuming that DM is the thermal relic of a new stable neutral particle with mass $M\sim T_{0}^{1/2}G_{\rm N}^{-1/4}
\sim\TeV$ where $T_{0}\sim 3\,{\rm K}$ is  the present temperature of the universe and
$G_{\rm N}$ is the Newton constant.
Attempts to address the Higgs mass hierarchy problem typically
introduce a rich amount of new physics at the weak scale, including DM candidates;
supersymmetry is widely considered as the most promising proposal~\cite{DM}.
However
(i) no new physics appeared so far at collider experiments:
the simplest solutions to the hierarchy problem start needing
uncomfortably high fine-tunings of the their unknown parameters~\cite{FT};
(ii) the presence of a number of unknown parameters (e.g.\ all sparticle masses)
obscures
the phenomenology of the DM candidates;
(iii) the stability of the DM candidates is the result of extra features introduced by hand
(e.g.\  matter parity).

We here explore an opposite, minimalistic approach:
focussing on the Dark Matter problem, we add to the Standard Model (SM)
extra multiplets ${\cal X}+\hbox{h.c.}$ with minimal spin, isospin and hypercharge quantum numbers,
and search for the assignments that provide most or all of the following properties:
\begin{enumerate}

\item The lightest component is automatically stable
on cosmological time-scales.

\item The only renormalizable interactions of ${\cal X}$ to other SM particles are of gauge type,
such that new physics is determined by one new parameter:  the tree-level mass $M$
of the Minimal Dark Matter (MDM) multiplet.

\item  Quantum corrections generate a mass splitting $\Delta M$
such that the lightest component of $\cal X$ is neutral.
We compute the value of $M$ for which the thermal relic abundance
equals the measured DM abundance.

\item The DM candidate is still allowed by DM searches.
\end{enumerate}
In section~\ref{list} we list the possible candidates.
In section~\ref{splitting} we compute the mass splitting.
In section~\ref{Omega} we compute the thermal relic  abundance of
${\cal X}$ and equate it to the observed DM abundance, inferring the DM mass $M$.
In section~\ref{DMexp} we discuss signals and constraints from DM experiments.
In section~\ref{coll} we discuss collider signals.
Section~\ref{concl} contains our conclusions and a summary of the results.

\begin{table}[t]
$$\begin{array}{|ccc|ccccc|}\hline
\multicolumn{3}{|c|}{\hbox{Quantum numbers}}
&\hbox{DM can}&\!\!\!\hbox{DM mass}&m_{{\rm DM}^\pm} - m_{\rm DM}\!\!\!&
\hbox{Events at LHC} & \hbox{$\sigma_{\rm SI}$ in} \\
 \SU(2)_L &\! {\rm U}(1)_Y\! &\hbox{Spin} &
 \hbox{decay into} & \hbox{in TeV} &\hbox{in MeV} & \hbox{$\int {\cal L}\,dt = $100/fb} &
 \hbox{$10^{-45}\,{\rm cm}^2$}\\ \hline
 \hline
2 & 1/2 & 0 & EL & 0.54 \pm 0.01 & 350 & 320\div510 & 0.2\\ 
2 & 1/2 & 1/2 & EH & 1.1 \pm 0.03 & 341 & 160\div330 & 0.2\\ 
\hline
3 & 0 & 0 & HH^* & 2.0 \pm 0.05 & 166 & 0.2\div1.0 & 1.3\\ 
3 & 0 & 1/2 & LH & 2.4 \pm 0.06 & 166 & 0.8\div4.0 & 1.3\\ 
3 & 1 & 0 & HH,LL & 1.6 \pm 0.04 & 540 & 3.0\div10 & 1.7\\ 
3 & 1 & 1/2 & LH & 1.8 \pm 0.05 & 525 & 27\div90 & 1.7\\ 
\hline
4 & 1/2 & 0 & HHH^* & 2.4 \pm 0.06 & 353 & 0.10\div0.6 & 1.6\\ 
4 & 1/2 & 1/2 & (LHH^*) & 2.4 \pm 0.06 & 347 & 5.3\div25 & 1.6\\ 
4 & 3/2 & 0 & HHH & 2.9 \pm 0.07 & 729 & 0.01\div0.10 & 7.5\\ 
4 & 3/2 & 1/2 & (LHH) & 2.6 \pm 0.07 & 712 & 1.7\div9.5 & 7.5\\ 
\hline
5 & 0 & 0 & (HHH^*H^*) & 5.0 \pm 0.1 & 166 & \ll1 & 12\\ 
5 & 0 & 1/2 & - & 4.4 \pm 0.1 & 166 & \ll1 & 12\\ 
\hline
7 & 0 & 0 & - & 8.5 \pm 0.2 & 166 & \ll1 & 46\\  \hline
\end{array}$$
\caption{\em\label{tab:1} {\bf Summary of the main properties of Minimal DM candidates}.
Quantum numbers are listed in the first 3 columns;
candidates with $Y\neq 0$ are allowed by direct DM searches only if appropriate non-minimalities are introduced.
The 4th column indicates dangerous decay modes,
that need to be suppressed (see sec.~\ref{list} for discussion).
The 5th column gives the DM mass
such that the thermal relic abundance equals the observed DM abundance (section~\ref{Omega}).
The 6th column gives the loop-induced mass splitting between neutral and charged DM components (section~\ref{splitting}); for scalar candidates a coupling with the Higgs can give a small extra contribution, that we neglect.
The 7th column gives the $3\sigma$ range for the number of events expected at LHC (section~\ref{coll}).
The last column gives the spin-independent cross section, assuming a sample vale $f=1/3$ for the uncertain nuclear matrix elements (section~\ref{DMexp}).
}
\end{table}

\section{The Minimal DM candidates}\label{list}
We consider the following extension of the SM:
\beq\Lag = \Lag_{\rm SM} +c
\left\{\begin{array}{ll}
 \bar{{\cal X}} (i D\hspace{-1.4ex}/\hspace{0.5ex}+M) {\cal X} & \hbox{when ${\cal X}$ is a spin 1/2 fermionic multiplet}\\
|D_\mu {\cal X}|^2 - M^2 |{\cal X}|^2& \hbox{when ${\cal X}$ is a spin 0 bosonic multiplet}
\end{array}\right.\eeq
where $D$ is the gauge-covariant derivative,
$c=1/2$ for a real scalar or a Majorana fermion
and $c=1$ for a complex scalar or a Dirac fermion:
in all cases we assign ${\cal X}$ in the minimal non-chiral representation
of the gauge group, and
$M$ is the tree-level mass of the particle.

\medskip

We want to identify the cases in which ${\cal X}$ provides a good DM candidate.
Therefore we assume the following gauge quantum numbers:
\begin{itemize}
\item[3.] ${\cal X}$ has no strong interactions~\cite{strong}.
\item[2.] ${\cal X}$ is an $n$-tuplet of the $\SU(2)_L$ gauge group, with $n=\{1,2,3,4,5,\ldots\}$.
\item[1.] For each value of $n$ there are few hypercharge assignments
that make one of the components of ${\cal X}$ neutral, $0=Q=T_3 + Y$
where $T_3$ is the usual `diagonal' generator of $\SU(2)_L$.
For a doublet, $n=2$, one needs $Y=1/2$.
For a triplet, $n=3$, one needs $Y=0$ (such that the component with $T_3 = 0$ is neutral),
or $Y=1$ (such that the components with $|T_3| = 1$ are neutral).
For a quadruplet, $n=4$, $Y=\{1/2,3/2\}$.
For a quintuplet, $n=5$, $Y=\{0,1,2\}$.
\end{itemize}
For each potentially successful assignment of quantum numbers
we list in table~\ref{tab:1} the main properties of the DM candidates.

The `decay' column lists the decay modes into SM particles that are
allowed by renormalizability, using a compact notation.
For instance, the scalar doublet in the first row
can couple as ${\cal X}_i L_{j\beta} E^\alpha \varepsilon_{ij}\varepsilon_{\alpha\beta}$
where $L$ is a SM lepton doublet, $E$ is the corresponding lepton singlet, $i,j$ are $\SU(2)_L$-indices, $\alpha,\beta$ are spinor indices,
and $\varepsilon$ is the permutation tensor; therefore the neutral component of ${\cal X}$ can decay as ${\cal X}_0 \to e\bar{e}$.
For another instance, the fermion doublet in the second row
can couple as ${\cal X}_{\alpha i} E_\beta H_j \varepsilon_{ij}\varepsilon_{\alpha\beta}$, where $H$ is the Higgs doublet: its neutral component can decay as ${\cal X}_0\to e h$.

In general, one expects also non-renormalizable couplings suppressed by $1/\Lambda^p$
(where $\Lambda$ is an unspecified heavy cut-off scale, possibly related
to GUT-scale or Planck-scale physics). These
give a typical lifetime
$\tau \sim \Lambda^{2p}\TeV^{-1-2p}$
for a particle with TeV-scale mass.
In order to make $\tau$ longer than the age of the universe\footnote{
We note that a $\tau$ comfortably longer than the age of the Universe already also prevents a decaying dark matter particle from having an impact on a number of cosmological and astrophysical observations (galaxy and cluster formation, type Ia supernovae, X--ray emissions from clusters, mass--to--light ratios in clusters, cosmic microwave background~\cite{decayDM,decayDM_CMB}).
E.g. measured CMB anisotropies constrain $\tau > 52$ Gyr at 95\% C.L..},
dimension-5 terms (i.e.\ $p=1$) must be effectively suppressed by $\Lambda \gg M_{\rm Pl}$,
while dimension-6 operators (i.e.\ $p=2$) are safe for $\Lambda\circa{>}10^{14}\GeV$.
Therefore in table~\ref{tab:1} we also list (in parenthesis)
the potentially dangerous dimension-5 operators.

\medskip

One sees that for low $n$ (upper rows of table~\ref{tab:1}) the multiplets
can interact with and decay into SM particles in a number of ways.
Actually, particles with these quantum numbers already appear
in a variety of different contexts: e.g.\
scalar triplets in little-Higgs models;
fermion or scalar triplet in see-saw models;
KK excitations of lepton doublets or of higgses in extra dimensional models;
higgsinos, sneutrinos, winos in supersymmetric models.\footnote{These DM candidates
typically behave very differently from the MDM candidates discussed in this paper.
However, these models have many free parameters, allowing a wide range of
different possibilities: in some corner of the parameter space  they can
reproduce the precise phenomenology of the corresponding MDM candidates.
See e.g.~\cite{Quasi-minimal,Antro,AM} for supersymmetric examples.}
In all these cases, a stable DM candidate can be obtained,
but only after suppressing the unwanted decay modes,
e.g.\ by invoking some extra symmetry.
All the best known WIMP DM candidates  need this feature
(e.g.\ matter parity in SUSY models~\cite{DM},
$T$-parity in little-Higgs models~\cite{Tparity},
KK-parity in `universal' extra dimension
 models~\cite{KK}\footnote{However, known ways of obtaining chiral fermions
 from extra dimensions introduce extra structures that
 generically break universality.}, etc.).
 In some cases the overall result of this model building activity looks plausible;
 in other cases it can look more like engineering rather than physics.

On the other hand, known massive stable particles (like the proton) do not decay for a simpler reason: decay modes consistent with renormalizability do not exist. DM can be stable for the same reason: for sufficiently high $n$ (lower rows of table~\ref{tab:1}), namely
\beq\hbox{$n\ge5$ for fermions, and $n\ge 7$ for scalars,}\eeq
${\cal X}$ is automatically stable because no SM particles have the quantum numbers
that allow sizable couplings to ${\cal X}$.
In other words, DM stability is explained by an `accidental symmetry', like proton stability.

\bigskip

An upper bound on $n$,
\beq\hbox{$n\le 5$ for Majorana fermions and $n\le 8$ for real scalars,}\eeq
is obtained by demanding perturbativity of
$\alpha_2^{-1}(E') = \alpha_2^{-1}(M) - (b_2/2\pi)\ln E'/M$ at $E'\sim M_{\rm Pl}$,
where $b_2 = -19/6 + c\ g_{\cal X} (n^2-1)/36$
with $c=1$ for fermions and $c=1/4$ for scalars.
Choosing a scale $E'$ much smaller than the Planck mass would not significantly
relax this upper bound on $n$, in view of the strong rise of $b_2$ with $n^3$,
and of the mild logarithmic dependence on $E'$.

Therefore table~\ref{tab:1} provides a realistically complete list of MDM candidates.

\section{The mass splitting}\label{splitting}
At tree level, all the components of ${\cal X}$ have the same mass $M$,
but loop corrections tend to make the charged components slightly heavier than the neutral one.

Actually, in the case of scalar DM there can be an extra source of mass splitting. The following extra renormalizable interactions can be present
\beq \Lag_{\rm non~minimal} = \Lag -c\, \lambda_H ({\cal X}^*  T^a_{\cal X} {\cal X})\, (H^* T^a_H H)-
c\, \lambda'_H |{\cal X}|^2 |H|^2 -   \label{eq:Lnonminimal}
 \frac{\lambda_{\cal X}}{2} ({\cal X}^*  T^a_{\cal X} {\cal X})^2 -  \frac{\lambda'_{\cal X}}{2} |{\cal X}|^4
 \eeq
 where $T^a_{R}$ are $\SU(2)_L$ generators in the representation to which ${R}$ belongs
 and $c\equiv 1~(1/2)$ for a complex (real) scalar.
 These extra interactions do not induce DM decay (because two ${\cal X}$ are involved, and we assume $\langle {\cal X}\rangle = 0$),
 and $\lambda_{\cal X},\lambda'_{\cal X}$ induce no significant effects.
 The coupling $\lambda_H$, however, splits the masses of the components of ${\cal X}$ by an amount
\beq \label{eq:tree}
\Delta M =\frac{ \lambda_H v^2 |\Delta T^3_{\cal X}|}{4M}  = \lambda_H \cdot 7.6\GeV \frac{\TeV}{M}\eeq
having inserted $\langle H\rangle  = (v,0)$ with $v=174\GeV$
and $\Delta T^3_{\cal X} = 1$.
In the following we assume that this possible correction to scalar masses is small enough that the
mass splitting is determined only by loop corrections.
We will see that this this is obtained for $\lambda_H \circa{<} {\mathcal O}(0.01)$, which is not a unreasonably strong restriction.

\medskip

The mass difference induced by loops of SM gauge bosons between two components of ${\cal X}$ with electric charges $Q$ and $Q'$ is explicitly computed as
\beq  M_Q -  M_{Q'} =\frac{\alpha_2 M}{4\pi}\left\{(Q^2-Q^{\prime 2})s_{\rm W}^2 f(\frac{M_Z}{M})+(Q-Q')(Q+Q'-2Y)
\bigg[f(\frac{M_W}{M})-f(\frac{M_Z}{M})\bigg]\right\}\eeq
where
\beq
f(r) =\left\{\begin{array}{ll}
+r  \left[2 r^3\ln r -2 r+(r^2-4)^{1/2} (r^2+2) \ln A\right]/2 & \hbox{for a fermion}\\
-r \left[2r^3\ln r-k  r+(r^2-4)^{3/2} \ln A\right] /4 & \hbox{for a scalar}
  \end{array}\right.\eeq   with $A = (r^2 -2 - r\sqrt{r^2-4})/2$ and $s_{\rm W}$ the sine of the weak angle.
   The constant $k$ is UV divergent, and can be reabsorbed in a renormalization of $M$ and of
$\lambda_H\to \lambda_H +4kY\alpha_2 \tan^2\theta_{\rm W}$;
notice that  if $Y=0$ the UV-divergent term $k$ produces no effect.
 We will be interested in the limit $M\gg M_{W,Z}$: for both scalars and fermions
  the result is well defined and equal to
$ f(r)\stackrel{r\to 0}{\simeq}  2\pi r$,
  such that
the mass splitting with respect to the neutral component, $Q'=0$, becomes:
\beq M_Q - M_0 \simeq Q(Q+ 2Y/\cos\theta_{\rm W})\Delta M\qquad\hbox{for $M\gg M_W,M_Z$}\eeq
where $\Delta M$ is the mass splitting between $Q=1$ and $Q'=0$ components
in the case of zero hypercharge, $Y=0$:
\beq \Delta M = \alpha_2 M_W \sin^2\frac{\theta_{\rm W}}{2}=(166\pm 1)\MeV.
\label{eq:166}\eeq
As we already observed, when $Y=0$  the next term in the  expansion of
$f(r)\stackrel{r\to 0}{\simeq} 2\pi r + (kr)^2$ gives no effect, such that
corrections due to finite $M$ are only of order $M_{W,Z}^2/M^2$,
and thereby practically negligible for the values of $M$ computed in the next section.
The values of $\Delta M$ for all the relevant MDM candidates are listed
   in table~\ref{tab:1}.

\smallskip

One can easily understand why such loop-induced
$\Delta M$ turns out to be not suppressed by $M$
(unlike the tree level term of eq.\eq{tree}), why
the neutral component gets lighter and why spin becomes irrelevant for $M\gg M_Z$.
The quantum correction $\Delta M$ is dominated by a well-known classical effect:
the Coulomb energy~\cite{Coulomb} stored in the electroweak
electric fields. Indeed a point-like charge $g$ at rest generates the electric potential
$\varphi(r) = g e^{-M_V r/\hbar}/4\pi r$
of one vector boson with mass $M_V$, such that its
Coulomb energy is (in natural units)
\beq\label{eq:Coulomb} \delta M = \int d^3r \left[\frac{1}{2}(\vec\nabla \varphi)^2 + \frac{M_V}{2}\varphi^2\right]=
\frac{\alpha}{2} M_V + \hbox{(UV-divergent constant)}.\eeq
The Coulomb energy on scales $M^{-1}\circa{<} r \circa{<} M_Z^{-1}$ gives
a common divergent renormalization of $M$.
The computable mass difference is the Coulomb energy on
scales larger than the $\SU(2)_L$-breaking scale 
$r/\hbar\circa{>}M_Z^{-1}$,
such that the  microscopic details of ${\cal X}$ on much smaller scales $\sim M^{-1}$ become irrelevant.
Inserting the $W,Z$ couplings eq.\eq{Coulomb} gives
$\Delta M = \alpha_2 (M_W - c_{\rm W}^2 M_Z)/2$ which is equivalent to eq.\eq{166}.

\section{The thermal relic MDM abundance}\label{Omega}
Assuming that DM arises as a thermal relic in the Early Universe, we can compute its abundance as a function of its mass $M$. Requiring that ${\cal X}$ makes all the observed DM, we can therefore univocally determine $M$.

We can neglect the mass splitting  $\Delta M \ll M$;
indeed eq.\eq{gamma} will confirm that
all DM components can be approximatively considered as stable
on time-scales comparable to the expansion rate at temperature
$T\sim M_W$.
The heavier charged components eventually decay into the lighter neutral DM,
before nucleo-synthesis and giving a negligible entropy release.
Moreover, following the common practice, we ignore thermal
corrections: we estimate that they mainly induce thermal
mass splittings of order $\Delta M\sim (g^2 T)^2/M$,
which likely can also be neglected.
A more careful investigation of this issue could be worthwhile.

Thanks to the above approximations we can write a single Boltzmann equation
that describes the evolution of the total abundance of all components ${\cal X}_i$ of the multiplet as a whole.
In particular, it includes all co-annihilations in the form of
 $\sum_{ij}\sigma_A({\cal X}_i {\cal X}_j \to {\rm SM~particles})$.
Since we will find that the observed DM abundance is obtained for $M^2 \gg M_Z^2$,
we can compute the relevant thermally-averaged cross-sections in the $\SU(2)_L$-symmetric limit.
Furthermore as usual the freeze-out temperature is $T_f\sim M/25\ll M$, such that
we can keep only the dominant $s$-wave (co)annihilation processes.
The final DM abundance  can be well approximated as~\cite{KT,altri}
\beq \frac{n_{\rm DM}(T)}{s(T)} \approx \sqrt{\frac{180}{\pi\ g_{\rm SM}}}\frac{1}{M_{\rm Pl}\ T_f \langle \sigma_A v\rangle},\qquad
\frac{M}{T_f} \approx \ln\frac{
g_{\cal X} M M_{\rm Pl}\langle\sigma_A v\rangle}{240\ g_{\rm SM}^{1/2}}\sim 26
\eeq
where $g_{\rm SM}$ is the number of SM degrees-of-freedom in thermal equilibrium
at the freeze-out temperature $T_f$, and $s$ is their total entropy.

If ${\cal X}$ is a scalar, annihilations into SM scalars and fermions are $p$-wave suppressed
(notice however that the large number of fermions present in the SM could partially compensate this suppression),
and the dominant annihilation channel is into $\SU(2)_L\otimes{\rm U}(1)_Y$ vector bosons: ${\cal X}{\cal X}^*\to AA$.
We find\footnote{We here only make use of
$${\rm Tr}\, T^a T^b = \delta^{ab}\frac{n}{12}(n^2-1),\qquad
{\rm Tr}\, T^a  T^a T^b T^b =\frac{n}{16}(n^2-1)^2
\qquad
{\rm Tr}\, T^a T^b T^a T^b =\frac{n}{16}(n^2-5)(n^2-1)$$
where $T^a$ are the  $\SU(2)_L$ generators
in the representation with dimension $n$.
The neutral DM component in the ${\cal X}$ multiplet
has vector-like NC gauge interactions with the $Z$ boson
and  vector-like CC interactions with the $W^\pm$ boson and with the charged $\DM^\mp$ component.
The gauge couplings are
$$ g_{\rm CC}^\pm=\frac{g}{\sqrt{2}} \cdot \frac{\sqrt{n^2-(1\mp 2Y)^2}}{2},\qquad
g_{\rm NC}=\frac{g\ Y}{c_{\rm W}}.$$}
\beq\label{eq:sigAs} \langle \sigma_A v\rangle \simeq\frac{g_2^4\ (3-4n^2+n^4) + 16\ Y^4 g_Y^4 + 8g_2^2 g_Y^2 Y^2 (n^2-1)}{64 \pi\ M^2\ g_{\cal X}}
\qquad\hbox{if ${\cal X}$ is a scalar}\eeq
where $g_{\cal X} = 2n$ for a complex scalar and $g_{\cal X}=n$ for a real scalar. 

%

\medskip

If ${\cal X}$ is a fermion, annihilations into gauge bosons are again described by  eq.\eq{sigAs}, where now
 $g_{\cal X}=4n$ for a Dirac fermion and $g_{\cal X}=2n$ for a Majorana fermion.
Furthermore, now also fermion and Higgs final states contribute to the $s$-wave cross section
of annihilations {\em plus} co-annihilations
(notice that when $Y=0$ only co-annihilations are present).
The result is
\beq \langle \sigma_A v\rangle \simeq
\frac{g_2^4\ (2n^4+17n^2-19) + 4  Y^2 g_Y^4 (41+8Y^2)+16g_2^2 g_Y^2 Y^2 (n^2-1)}{128\pi\ M^2\ g_{\cal X}}
\qquad\hbox{if ${\cal X}$ is a fermion}.
\eeq
Table~\ref{tab:1} shows the values of $M$ needed to reproduce all the observed DM.\footnote{
As discussed in later works (\hepart[hep-ph/0610249]{J. Hisano et al.} and
\hepart[0706.4071]{M. Cirelli et al.}) non perturbative corrections affect the
cosmological density in an important way with respect to the perturbative result obtained here.}
Within the standard $\Lambda$CDM cosmological model, present data demand
 $\Omega_{\rm DM} h^2=0.110\pm 0.006$ i.e.\
 $n_{\rm DM}/s = (0.40\pm0.02)\eV/M$~\cite{WMAP}.
In all cases it turns out that $M^2 \gg M_{W,Z}^2$, justifying our approximation of neglecting $\SU(2)_L$-breaking
corrections to $\langle \sigma_A v\rangle$.
Since $\Omega_{\rm DM}\propto M^2$, the values of $M$ reported in table~\ref{tab:1} suffer a $3\%$ experimental uncertainty.
We report numerical values indicating only the experimental uncertainty
because it is difficult to quantify the  theoretical uncertainty:
it can give a comparable contribution, and can be reduced by performing refined computations.

\medskip

In the scalar case, the non-minimal couplings in eq.\eq{Lnonminimal} are generically allowed and give extra annihilations into higgses:
\beq 
\label{eq:extra}
\langle \sigma_A v\rangle_{{\rm extra}} = 
\frac{|\lambda_H^{\prime}|^2 + (n^2-1) |\lambda_H|^2/16}{16\pi\ M^2\ g_{\cal X} }
\qquad\hbox{if ${\cal X}$ is a non-minimal scalar}.
\eeq
(no interference terms are present).
The contribution from $\lambda_H$ is negligible for the values that allow to neglect its effect on mass splittings (discussed in eq.\eq{tree}), and the contribution  from $\lambda'_H$ is negligible
for $|\lambda'_H| \ll g_{Y}^2,g_2^2$. We assume that this is the case so that these extra terms are negligible for all candidates with electroweak interactions.\footnote{However, when ${\cal X}$ is a neutral scalar singlet,  these non-minimal annihilations are the only existing ones: the observed amount of DM is obtained for $M\approx 2.2\TeV|\lambda'_H|$ (we are assuming $M\gg M_Z$; for generic values of $M$ the correlation between  $M$ and $\lambda'_H$ was studied in~\cite{Burgess}). We stick to the minimal setup so that the singlet is not a useful candidate for our purposes. This holds a fortiori for the case of fermionic DM.}
In this paper we assume that this is the case, and here study this issue quantitatively: to compute how much the inferred value of $M$ would be affected one simply adds the contribution in eq.\eq{extra} to those in eq.\eq{sigAs}: the mass of the low $n$ candidates are more affected than the high $n$ ones, because the gauge-mediated annihilation cross sections are relatively less important for the formers. 
For example, a large $\lambda'_H = 1$ would increase the predicted value of
$M$ by a factor $2.4$ for $n=2$, by $20\%$ for $n=3$, by $2\%$ for $n=5$,
and by $0.5\%$ for $n=7$.
Note that $\lambda_H\sim g_2^2$ and $\lambda'_H\sim g_Y^2$ are the values
predicted by some solutions to the hierarchy problem, such as supersymmetry and gauge/higgs unification.


\medskip

A smaller $M$ is allowed if ${\cal X}$ gives only a fraction of the total DM density.
In the more general situation where many MDM multiplets are present,
to a good approximation their abundances evolve independently,
such that the observed DM abundance is reproduced for lower values of DM masses.
In this more general situation, the $M$ values in table~\ref{tab:1} must be reinterpreted as
upper bounds on $M$.
For example, with $N$ identical families of the same ${\cal X}$ particle
their common mass becomes $N^{1/2}$ times lower.

\medskip

Finally, we notice that the combination of
a precise collider measurement of $M$ with
a precise measurement and computation of the DM abundance,
could allow to test the expansion rate at a freeze-out temperature around
the electroweak scale.
This is a particularly interesting region:
one can test if the electroweak vacuum energy
behaves just like as a fine-tuned constant,
or if some sort of relaxation mechanism of the cosmological constant takes place.

\section{Low-energy MDM signals}\label{DMexp}
\subsubsection*{Direct detection}
Searches for elastic DM collision on nuclei ${\cal N}$ provide strong constraints.
MDM candidates with $Y\neq 0$ have vector-like interactions with the $Z$ boson that produce
spin-independent elastic cross sections 
\beq \sigma({\rm DM} \, {\cal N}\to {\rm DM}\,{\cal N}) =c\frac{G_{\rm F}^2M_{\cal N}^2}{2\pi} Y^2(N - (1-4s_{\rm W}^2) Z)^2 \eeq
where $c=1$ for fermionic DM and $c=4$ for scalar DM~\cite{GoodWit};
$Z$ and $N$ are the number of protons and of neutrons in the target nucleus with mass $M_{\cal N}$, 
We are assuming $M\gg M_{\cal N}$.
For all DM candidates with $Y\neq 0$, this elastic cross section is
 $2\div 3$ orders of magnitude above present bounds~\cite{CDMS}.
 Such MDM candidates are therefore excluded, unless
 minimality is abandoned in appropriate ways that allow
 to avoid the experimental limit~\cite{Quasi-minimal,Antro};
 this for instance happens is the well-known case of the Higgsino: the mixing
 with Majorana gauginos can split its components by an amount $\delta m$
such that the lightest one becomes a Majorana fermion which cannot have a vector-like
coupling to the $Z$ boson.
In our case, if a $\delta m$ larger than the DM kinetic energy and smaller than $M_Q-M_0$ is generated by some similar non-minimal mechanism, it would affect only the direct detection DM signals  such that all candidates with $Y\neq 0$
become still allowed: therefore they are included in table~\ref{tab:1}.
However, in the following we insist on minimality and focus mainly on the candidates with $Y=0$.

MDM candidates with $Y=0$ have vanishing $\DM \,{\cal N}$ and $\DM\, \DM$
cross sections at tree level, but are accompanied by a charged component heavier by $\Delta M=166\MeV$ (see eq.\eq{166}),
that produces various possible effects.

\bigskip

\begin{figure}[t]
$$
\includegraphics[width=\textwidth]{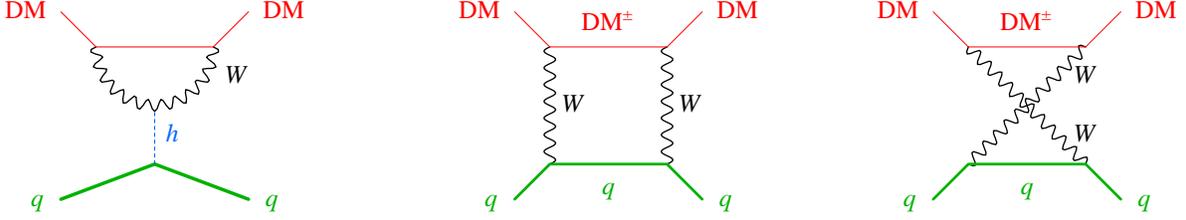}$$
\caption{\label{fig:FeynLoop}\em One loop DM/quark scattering for fermionic MDM with $Y=0$.
Two extra graphs involving the four particle vertex exist in the case of scalar MDM.}
\end{figure}

An elastic cross section on nuclei is generated at loop level via the diagrams in fig.\fig{FeynLoop}.
An explicit  computation of the relevant one-loop diagrams is needed
to understand qualitatively and quantitatively the result.
We find that non-relativistic MDM/quark interactions
of fermionic MDM with mass $M\gg M_W \gg m_q$
are described by the effective Lagrangian
\beq\label{eq:directW} \Lag_{\rm eff}^W =(n^2-(1\pm 2Y)^2)\frac{\pi \alpha_2^2}{16 M_W}\sum_q
\left[
(\frac{1}{M_W^2}+\frac{1}{m_h^2})  [\bar {\cal X}{\cal X}] m_q [\bar{q}{q}]-\frac{2}{3M}
{ [\bar {\cal X}\gamma_\mu\gamma_5 {\cal X}][\bar {q}\gamma_\mu\gamma_5 {q}] }\right]\eeq
where the $+$ ($-$) sign holds for down-type (up-type) quarks $q=\{u,d,s,c,b,t\}$, 
$m_h $ is the Higgs mass and $m_q$ are the quark masses.
The first operator gives dominant spin-independent effects and is not suppressed by $M$;
the second operator is suppressed by one power of $M$ and gives spin-dependent effects.
Parameterizing the nucleonic matrix element as
\beq \langle{N} | \sum_q m_q \bar{q}q | N\rangle\equiv  f m_N  \eeq
where $m_N$ is the nucleon mass, the spin-independent DM cross section on a target nucleus ${\cal N}$ with mass $M_{\cal N}$
is given by
\beq\label{eq:sigmaSI} \sigma_{\rm SI}(\DM\,{\cal N}\to \DM\,{\cal N})
=(n^2-1)^2 \frac{\pi\alpha_2^4 M_{\cal N}^4   f^2}{64 M_W^2}(\frac{1}{M_W^2}+\frac{1}{m_h^2})^2.
\eeq

\begin{figure}[t]
$$\hspace{-5mm}
\includegraphics[width=0.8\textwidth]{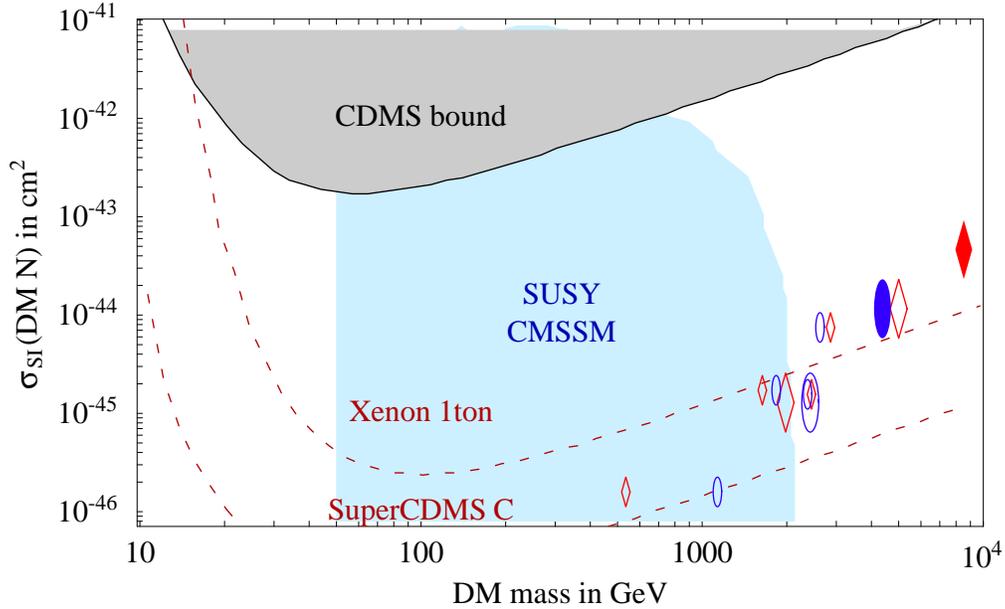}$$
\caption{\label{fig:direct}\em
{\bf Spin-independent cross sections per nucleon of MDM candidates} assuming the matrix element $f=1/3$.
Fermionic MDM candidates are denoted as blue circles; scalars as red diamonds.
The fully successful candidates (automatically stable $Y=0$ multiplets with $n=5,7$) are represented by filled symbols. We also show, as large empty symbols, the $Y=0$ candidates that require a stabilization mechanism and, with smaller symbols, those with $Y\neq 0$ (viable if an additional mechanism forbids a  much larger cross section mediated by a $Z$ boson).
The dashed lines indicate the sensitivity of some future experiments~\cite{future}.
The cloud indicates the range of values favoured by a minimal SUSY model.
}
\end{figure}

In the case of scalar MDM we find in the relevant non-relativistic limit:
an $M$-independent contribution to $\sigma_{\rm SI}$ equal to
the fermionic result of eq.\eq{sigmaSI};
an UV-divergent effect suppressed by $M$ that corresponds to a renormalization of
$|{\cal X}|^2|H|^2$ operators
(that can produce a much larger $\sigma_{\rm SI}$ if present at tree level);
no spin-dependent effect.

Assuming $m_h=115\GeV$ and
$f\approx 1/3$\footnote{To properly compute nuclear matrix elements
one must keep quarks off-shell, finding several operators that become
equivalent on-shell~\cite{Drees}:
$$ m_q[\bar {\cal X}{\cal X}]  [\bar{q}q],\qquad
[\bar {\cal X}{\cal X}] [\bar q i\ds q],\qquad
\frac{4}{3M}[\bar {\cal X}{i\partial_\mu}\gamma_\nu{\cal X}][\bar q i(\partial_\mu
\gamma_\nu + \partial_\nu\gamma_\mu - \frac{\eta_{\mu\nu}}{2}\ds)q],\ldots$$
Summing over all quarks the matrix elements are
$f\approx (0.3 \div 0.6)$ \cite{coupl} for the first operator,
  $f\approx 1.2$ for the third operator,
  while the matrix element of the second operator is unknown.
  In our computation, only the first operator contributes to the SI effects suppressed by the higgs mass,
 while the other SI effects arise from a combination of the various operators in proportion $0:-1:2$.
 Therefore cancellations are possible.
 We do not fully agree with result of a previous computation~\cite{HisanoDirect},
 performed for the fermionic
 supersymmetric DM candidates: wino ($n=3$, $Y=0$) and Higgsino  ($n=2$, $|Y|=1/2$).}
(QCD uncertainties induce one order of magnitude uncertainty on $\sigma_{\rm SI}$)
we plot in fig.\fig{direct} the MDM prediction for the standard nucleonic~\cite{DM,Drees,Pittel} cross section parameter, showing that it requires experimental sensitivities about 3 orders of magnitude larger than the current ones, but within the sensitivity of experiments currently under study.
The annual modulation effect of the DAMA/NaI experiment~\cite{dama} cannot be explained by MDM candidates, since they have too large masses and too small cross sections with respect to the properties of a WIMP compatible with the effect. 
We also show the signal rates for the MDM candidates with $Y\neq 0$,
assuming that they are resurrected by splitting their  neutral components.
In this case, the $W$ contribution of eq.\eq{directW}
to the effective Lagrangian relevant for direct DM signals becomes
2 times lower, and there is an extra $Z$ contribution given on-shell by
$$ \Lag_{\rm eff}^Z = \frac{{ \alpha_2^2\pi Y^2 }}{c_{\rm W}^4 M_Z}
\sum_q \left[3
[\bar {\cal X}{\cal X}] m_q [\bar{q}{q}] \left(\frac{(g_{Lq}+g_{Rq})^2}{2M_Z^2}+
\frac{1}{4m_h^2}\right)-
\frac{[\bar {\cal X}\gamma_\mu\gamma_5 {\cal X}][\bar {q}\gamma_\mu\gamma_5 {q}] }{M}(g_{Lq}^2 + g_{Rq}^2)
\right]
$$
where
$g_L = T_3 - Qs_{\rm W}^2$ and $g_R = -Qs_{\rm W}^2$ describe $Z$/quark couplings.

For comparison, the cloud in fig.\fig{direct} represents a typical range of values
found in scannings over the MSSM parameter space~\cite{SUSYscan},
restricted by present bounds and by naturalness assumptions.

\subsubsection*{CC direct detection?}
MDM could be probed via the tree level CC process $\DM \, {\cal N} \to {\cal N}^\pm \DM^\mp$.
In ordinary conditions this process is kinematically forbidden, because
neither the DM kinetic energy, nor the ${\cal N}$ kinetic energy,
nor the  nuclear mass difference is large enough
provide an energy larger than the $\Delta M$ of eq.\eq{166}.
In these conditions a virtual $\DM^\pm$ can be produced slightly off-shell,
but the cross section for the resulting
multi-body process $\DM\, {\cal N}\to \DM\,{\cal N}^\pm\, e^\mp\nubarnu_e $
 is negligibly small.
Production of on-shell $\DM^\pm$ could take place in environments,
like cluster of galaxies or supernov\ae,
where DM or nuclei have enough kinetic energy;
but we do not see how this can provide a detectable signal.



\medskip

We here explore the possibility of accelerating nuclei in a controlled experiment:
a dedicated discussion seems worthwhile because prospects of searching for
this unusual signal seem not much beyond  what can be considered as realistic.
Protons $p$ or nuclei ${\cal N}$ can be accelerated up to energy $E_p > \Delta M$
(while collider experiments need a much larger energy $E_p\circa{>} M$)
and stored in an accumulator ring.
The DM background acts as a diffuse target such that there is no need for focusing the beam.
The cross section is more easily computed in the limit $\Delta M,m_p \ll E_p \ll M$.
Averaging  over initial colors and polarizations we find the following
partonic cross section that applies for both fermionic and scalar MDM:
\beq\label{eq:DMp} \hat\sigma(a\,{\rm DM}\to a' \,{\rm DM}^\pm)=\sigma_0\frac{n^2-1}{4}
 \bigg[1-\frac{\ln (1+4E^2/M_W^2)}{4E^2/M_W^2}\bigg],\qquad
\sigma_0=\frac{G_{\rm F}^2 M_W^2}{\pi}=1.1\,10^{-34}\,{\rm cm}^2\eeq
where $a$ is any quark or anti-quark (or electron) with energy $E$.
Therefore the $p\,\DM\to n\,\DM^+$ cross section approaches the constant value $3\sigma_0(n^2-1)/4$
for  $M_W\circa{<} E_p\circa{<} M$.
Increasing the proton energy $E_p$ above $M_W$  increases the $\DM^\pm$ velocity,
$\beta\sim \max (10^{-3},(E_p/M)^{1/2})$,
but does not increase the cross section.
Such a high cross section, not suppressed by powers of $M$,
arises because this is a non-abelian Coulomb-like process.
For comparison, SUSY DM candidates have comparable cross sections
(lower if the neutralino is dominantly bino) and
chargino/neutralino mass splittings $\Delta M$ possibly in the 10 GeV range.\footnote{A better experiment might be possible in the case of supersymmetric neutralino DM, but only after that
sparticle masses will be precisely measured:  an electron beam can be stored
at the energy that makes the process electron~neutralino $\to$ selectron resonant,
with the production of electrons of known energy~\cite{HisanoAgain}.
In this respect, we also note that plans for neutrino factories are independently proposing muon beams (less plagued by synchrotron losses than electron beams) that could have the desired high intensity ($10^{21}$ muons stored per year?), time-scale (2020?) and energy (few tens of GeV?) for a muon~neutralino $\to$ smuon experiment; for detection, a muon signal is better than an electron signal.}

At energies $E_p\ll M_W$ the cross section decreases as $(E_p/M_W)^2$,
and when $E_p\circa{<} m_p$ one should switch from a partonic to a nucleonic description.
Nuclear cross sections are not coherently enhanced for the $\Delta M$ values suggested by MDM.
The event rate is
\beq \frac{dN}{dt}=\varepsilon N_p \sigma \frac{\rho_{\rm DM}}{M}=\varepsilon
\frac{10}{{\rm year}} \frac{N_p}{10^{20}} \frac{\rho_{\rm DM}}{0.3\GeV/{\rm cm}^3}\frac{\TeV}{M}
\frac{\sigma}{3\sigma_0}
\eeq
where $\rho_{\rm DM}$ is the local DM density and
$\varepsilon$ is the detection efficiency,
related e.g.\ to the fraction of beam that can be monitored.
A circular accumulator ring that employs a magnetic field $B$ has radius
$R = E_p/q_p B = 21\,{\rm m} (E_p/10\GeV)(10\,{\rm Tesla}/B)$.
Proton drivers currently planned for neutrino beam experiments can produce
more than $10^{16}$ protons per second,
and accumulating $N_p\sim 10^{20}$ protons is considered as possible.

The main problem seems disentangling the signal from the beam-related backgrounds.
This issue crucially depends on how $\DM^\pm$ behaves;
e.g.\ an enough long lived $\DM^\pm$ decaying into $\mu^\pm$ would give a clean enough signal.
In section~\ref{coll} we compute the $\DM^+$ life-time, finding that it is long lived, but not enough.


\subsubsection*{Indirect detection}

DM DM annihilations can have larger cross-sections.
Scalar DM annihilations into higgses can arise at tree level.
Furthermore, as noticed in the context of
SUSY models with multi-TeV LSP mass\footnote{In this corner of
its parameter space SUSY does not naturally provide
electroweak-symmetry breaking:
cancellations among individual corrections
to the squared Higgs mass which are thousand of times larger than their sum are needed.
In this region DM DM annihilations are negligibly
affected by all other sparticles outside the DM multiplet,
restricted to be higgsino ($n=2$) or wino ($n=3$).}
annihilations into gauge bosons of DM particles
with non relativistic  velocity $\beta\circa{<}\alpha_{\rm em}$
(in our galaxy $\beta\sim 10^{-3}$)
can be strongly enhanced
by the presence of quasi-degenerate charged $\SU(2)_L$ partners~\cite{Hisano}.
For MDM candidates with $Y=0$, by performing a
na\"{\i}ve perturbative computation of the relevant Feynman diagrams
(see~\cite{preHisano} for analogous computations in the context of supersymmetry), we find
\begin{eqnsystem}{sys:sss} \label{eq:20a}
\sigma(\DM\,\DM\to W^+ W^-) v &=& (n^2-1)^2\frac{\pi \alpha_2^2}{32M^2},\\
\riga{at tree level, and}\\[-5mm]
\sigma(\DM\,\DM\to \gamma\gamma) v &=& (n^2-1)^2\frac{\pi \alpha^2_{\rm em}\alpha_2^2}{16 M^2_W}
\end{eqnsystem}
at one loop.  Analogous results hold for $Z\gamma$ and $ZZ$ final states.
However, the $\DM^0\DM^0$ system can happen to have mass equal to the
$\DM^+\DM^-$ or $\DM^{++}\DM^{--}$ systems:
the mass difference $\Delta M \sim \alpha_2 M_W$ can be compensated by the binding energy
$E_B\sim \alpha_2^2 M$ of the two-body state.
This happens for specific values $M_{*}\sim M_W/\alpha_2$ of $M$:
if $M\approx M_*$ the cross sections $\sigma(\DM\,\DM \to AA)$ get enhanced by a factor
${\cal O}(1-M/M_*)^{-2}$ and acquire comparable values for all vector bosons $A$~\cite{Hisano}.
Unlike in the supersymmetric case,
in the MDM case $\Delta M$ and consequently $M_*$ are univocally predicted.
We find resonances at\footnote{We here outline the computation,
performed along the lines of~\cite{NR,Hisano}.
Using the case $n=5$ as an example,
the Hamiltonian that describes the non-relativistic  canonically normalized
$\{\DM^{++}\DM^{--}$, $\DM^+\DM^-$, $\DM^0\DM^0\}$
two-body states is
$$ H  =- \frac{\nabla^2}{M}+
\Delta M\diag(8,2,0)-\frac{1}{r}
\pmatrix{4\alpha_{\rm em} + 4\alpha_2 c_{\rm W}^2 e^{-M_Z r}& 2 \alpha_2 e^{-M_Wr} &0\cr
2\alpha_2 e^{-M_Wr}&
 \alpha_{\rm em} + \alpha_2 c_{\rm W}^2 e^{-M_Z r} &
3\sqrt{2} \alpha_2 e^{-M_Wr}\cr
0&3\sqrt{2} \alpha_2 e^{-M_Wr}& 0}.
$$
One recognizes the kinetic energy in the center-of-mass frame,
the contributions due to the mass difference $\Delta M$, the Coulomb
energy $Q^2\alpha_{\rm em}/r$ and its non-abelian generalizations.
We numerically found the values of $M$ that give rise to $s$-wave eigenstates with $H=0$,
relevant for $\DM^0\DM^0$ annihilations.
}
\beq\label{eq:resonant}\begin{array}{c|ccccc}
n & \multicolumn{5}{|c}{M_* \hbox{ in TeV}}\\ \hline
3 & 2.5 & 9.8&\ldots\\
5 &1.8&3.3&6.6&\ldots\\
7 &.74&1.6&2.9&3.7&\ldots
\end{array}\eeq
In the case $n=3$ the first resonances happens to be close to the value of $M$
suggested by the DM abundance, giving a ${\cal O}(100)$ enhancement.

\bigskip

DM DM annihilations can
produce different types of signals
relevant for indirect  DM searches.
We do not perform more precise  computations of the relevant particle-physics
because the predicted rates are significantly affected by astrophysical uncertainties
concerning the DM distribution in our Galaxy.

When DM annihilations occurs in the inner core of the Earth or the Sun,
producing a neutrino flux to be searched at neutrino telescopes, the
signal is strictly linked to the situation which occurs for direct
detection. In this case the signal depends on the scattering cross
section on the nuclei of the capturing body (Earth or Sun). We
estimate that the MDM candidates  compatible with direct detection
(plotted in fig.\ \ref{fig:direct}) provide a neutrino
signal which is about two orders of magnitude lower than current
sensitivities \cite{amanda}. Large area detectors, like e.g.\ ICECUBE
\cite{icecube}, might access the upgoing muon signal for these DM
candidates.
The neutrino energy and flavour spectra can be computed
in terms of particle physics along the lines of~\cite{DMnu}.

In the case of DM annihilation inside the Galactic halo, the signals
consist in fluxes of antiprotons, antideuterons, positrons and
gamma--rays. We estimate that an annihilation cross section around
$10^{-23}$ cm$^{3}$/s is needed in order to provide an antiproton
signal which can emerge over the expected background for antiproton
kinetic energies lager than about 100 GeV, and without falling in
conflict with the lower energy data~\cite{pbar1}.  The off-resonance
cross section of eq.\eq{20a} can only reach $10^{-(25\div 26)}$
cm$^{3}$/s for $n=\{5,7\}$.  Signals at the level of the expected
background can be approached for DM masses that lie within $\sim 10\%$
of the resonant values $M_\star$ discussed above. In this case,
forthcoming experiments like PAMELA \cite{pamela} and AMS~\cite{AMS}
have a chance to observe a signal. However, we recall that the
antiproton signal is affected by a large astrophysical uncertainty, of
the order of a factor of a few \cite{pbar1}. This may
induce a more favourable situation for the antiproton signal,
especially in the case of a thick confinement region in the Galaxy.
As for positrons, again enhanced cross sections for $M$ close to the
resonant values $M_\star$ can provide signals at the level of
detectability \cite{Hisano}.

A cross section at the level of providing a signal in the antiproton
channel would also give a antideuteron signal \cite{dbar0} comparable
to its background for energies larger than about 50 GeV.  However,
planned experiments like GAPS~\cite{GAPS} are sensitive to the more
relevant low-energy tail \cite{dbar0}; DM annihilations into gauge
bosons are not a favourable channel for antimatter
production~\cite{pbar1, profumo}.  Therefore, the possibility to
detect antideuterons from MDM looks rather difficult.

Finally, in the case of gamma-rays, we extimate that a cross section
at the level of $ 10^{-24}$ cm$^{3}$/s can allow detection of
a signal from the galactic center for the GLAST experiment~\cite{GLAST,pieri}, in the
energy range around 100 GeV and
assuming a Navarro-Frenk-White density profile.
Cross sections one order of magnitude smaller
can be accessed by future water \v{C}erenkov detectors like
VERITAS~\cite{VERITAS,pieri}. Again, also in this case only
a resonantly enhanced
cross section can provide a detectable signal; in such a case
a signal coming from external Galaxies like M31 could
be accessed from experiments like EGRET and VERITAS \cite{pieri}.

\section{High energy MDM signals}\label{coll}
MDM multiplets give one loop `universal' corrections to
electroweak precision data affecting only the $W,Y$ parameters~\cite{STWY} as 
\beq
W = c\ g_{\cal X} \frac{\alpha_2}{60\pi} \frac{M_W^2}{M^2} \frac{n^2-1}{12},\qquad
Y =c\ g_{\cal X} Y^2\frac{\alpha_Y}{60\pi}  \frac{M_W^2}{M^2}\eeq
where $c=1$ for fermionic DM and $c=1/4$ for scalar DM.
For the masses $M$ computed in section~\ref{Omega} one gets a negligibly small
$W,Y\sim 10^{-7}$.

\medskip

Direct pair production of MDM particles is a more promising future signal.
We focus on the LHC collider, that will collide $pp$ at $\sqrt{s} = 14 \TeV$.
For $Y=0$ the partonic total cross sections
(averaged over initial colors and spins)
for producing all DM components are
\beq \hat\sigma_{u\bar d}= \hat\sigma_{d\bar u}=2\hat\sigma_{u\bar u} = 2\hat\sigma_{d\bar d}=
\frac{g_{\cal X} g_2^4(n^2-1)}{13824\ \pi \hat s}\beta \cdot\left\{\begin{array}{ll}
\beta^2 & \hbox{if ${\cal X}$ is a scalar}\\
3-\beta^2&\hbox{if ${\cal X}$ is a fermion}\end{array}\right.\eeq
where  the subscripts denote the colliding partons,
and $\beta = \sqrt{1-4M^2/\hat{s}}$ is the  DM velocity with respect to
the partonic center of mass frame.
Production of non-relativistic scalars is $p$-wave suppressed in the usual way.
In table~\ref{tab:1} we show the number of $pp\to\DM_i\,\DM_j \,X$
 events (here $X$ denotes any other particles). We computed cross sections including the hypercharge contribution,
summed over all DM components,
 and assumed an integrated luminosity equal to
 $100/{\rm fb}$, that each detector will accumulate in
 $10^7\,{\rm sec}$ (i.e.\ one collider-year)
if LHC will deliver its planned luminosity, ${\cal L}\approx 10^{30}/{\rm cm}^2{\rm sec}$.
The event rate depends significantly on the precise value of the DM mass $M$,
but surely some MDM candidates are too heavy for LHC.
A collider with beam energy $2\div 4$ times higher could fully test all MDM candidates.
Possible upgrades of LHC luminosity and magnets are discussed in~\cite{LHC28}.


\medskip

MDM has a clean signature, that allows discovery even if only a few events are seen.
The small mass splitting among the DM components makes
too hard to tag the missing energy carried away by neutral DM particles,
but also makes charged MDM component(s) enough long-lived that they
manifest  in the detector as charged tracks.
Irrespectively of the DM spin the life-time of $\DM^\pm$ particles with $Y=0$
and $n=\{3,5,7,\ldots\}$ is
$\tau \simeq 44\,{\rm cm}/(n^2-1)$
and the decay channels are
\beq\label{eq:gamma}
\begin{array}{lcll}
\DM^\pm \to \DM^0 \pi^\pm \qquad&:&\displaystyle
\Gamma_\pi =(n^2-1) \frac{G_{\rm F}^2V_{ud}^2\, \Delta M^3 f_\pi^2}{4\pi}
\sqrt{1-\frac{m_\pi^2}{\Delta M^2}},\qquad
& {\rm BR}_\pi =97.7\%\\[3mm]
\DM^\pm \to \DM^0 e^\pm\nubarnu_e &:& \displaystyle
\Gamma_e = (n^2-1) \frac{G_{\rm F}^2 \,\Delta M^5}{60\pi^3} &
{\rm BR}_e=2.05\%\\[3mm]
\DM^\pm \to \DM^0 \mu^\pm\nubarnu_\mu &:& \displaystyle
\Gamma_\mu =0.12\ \Gamma_e &
{\rm BR}_\mu=0.25\%\\
\end{array}
  \eeq
 having used the normalization $f_\pi = 131\,{\rm MeV}$~\cite{CHPT}
 and the $\Delta M$ of eq.\eq{166},
 which accidentally happens to be the value that maximizes ${\rm BR}_\pi$.
The $\DM^+$ life-time is long enough that decays can happen inside the detector.
  On the contrary, the faster decays of $\DM^{\pm\pm}$ particles (present for $n\ge 5$)
  mostly happen within the non-instrumented region
with few cm size around the collision region.
Measurements of $\tau$ and of the
energy of secondary soft pions, electrons and muons
constitute tests of the model, as these observables negligibly depend on the DM mass $M$.
On the contrary, measurements of the total number of events or of the
DM velocity distribution would allow to infer its mass $M$ and its spin.
Although SM backgrounds do not fully mimic the well defined MDM signal,
at an hadron collider (such as the LHC) their rate is so high that
some experimentalists consider impossible triggering on the MDM signal.

%

Notice that extra $\SU(2)_L$ multiplets that couple (almost) only through gauge interactions
tend to give a LHC phenomenology similar to the one discussed above,
irrespectively of their possible relevance for the MDM problem.
On the contrary, DM candidates like neutralinos are often dominantly produced
through gluino decays, such that
DM is accompanied by energetic jets rather than by charged tracks.

\section{Conclusions}\label{concl}
We extended the Standard Model by adding a spin-0 or spin-1/2 $n$-tuplet of $\SU(2)_L$ with hypercharge $Y$
that only has gauge interactions and mass $M$.
Some multiplets contain neutral components, that are potential Dark Matter (DM)  candidates.
\begin{itemize}
\item Multiplets with $Y \neq 0$ are already excluded by direct DM searches.
They can be resurrected by introducing non-minimal mechanisms that prevent  $Z$-mediated
 DM/nuclei coupling, e.g.\ by appropriately mixing their
neutral components with a  singlet.
\item Multiplets with $Y=0$ and odd $n=\{3,5,7,\ldots\}$  contain allowed DM candidates.
\begin{itemize}
\item For $n=3$ one needs to impose DM stability by hand.
\item For $n\ge 5$ the stability is instead automatically guaranteed by renormalizability, much alike proton stability.
\end{itemize}\end{itemize}
The set of interesting candidates is bounded by $n\circa{<}7$ in order to avoid Landau poles in $\alpha_2$.

Gauge interactions are spontaneously broken and thereby induce a non-trivial and peculiar Minimal DM (MDM) phenomenology,
fully computable in terms of a single unknown parameter: the DM mass $M$.
Electroweak breaking effects induce a mass splitting $\Delta M \sim \alpha M_W$
among the components of any given multiplet,
making the neutral component lighter than the charged components.
Assuming that only one MDM multiplet is present, its mass $M$ is determined by the request that  its relic thermal abundance equals the observed DM abundance.
Co-annihilations  play a crucial r\^ole, giving $M\sim $ few TeV.
Since $M\gg M_W$, various MDM properties depend dominantly only on the MDM gauge charge,
while the microscopic MDM properties (such as their spin) become irrelevant.

\medskip

The simplest fully successful MDM candidate  is a fermionic $\SU(2)_L$ quintuplet with mass $M\approx 4.4\TeV$.
MDM candidates are listed in table~\ref{tab:1}:
some are fully successful (automatically stable and consistent with DM searches),
others require
a stabilization mechanism (e.g.\ the wino-like candidate)
or a way to elude the bounds from direct DM searches (we list only those with $n\le 4$),
or both (e.g.\ the Higgsino-like candidate).
If multiple MDM multiplets exist,
all their masses become lighter than in table~\ref{tab:1}:
e.g.\ $42\%$ lighter in presence of 3 identical families of a single multiplet,
significantly increasing the number of events expected at LHC.

MDM multiplets contain charged components, slightly heavier than the neutral DM component.
For $Y=0$ the charged $\DM^\pm$ is $\Delta M = 166\MeV$ heavier
and has a life-time $\tau =44\,{\rm cm}/(n^2-1)$, giving a clean displaced-vertex signature at colliders.
The branching ratios are predicted to be
$\hbox{BR}(\DM^\pm \to \pi^\pm\DM^0)= 97.7\%$,
$\hbox{BR}(\DM^\pm\to\DM^0 e^\pm\nubarnu_e) = 2.05\%$,
$\hbox{BR}(\DM^\pm\to\DM^0\mu^\pm\nubarnu_\mu) = 0.25\%$.
We computed the event rate at LHC, finding that
LHC cannot probe all MDM multiplets, being more sensitive
to the ones with lower $n$ (and, if multiple multiplets are present,
to the ones that would give subdominant contributions to the DM density).

On the contrary direct DM searches are more sensitive
to higher $n$ MDM multiplets (and to ones that dominate the DM density).
Indeed one-loop diagrams generate a
  spin-independent MDM/nucleus cross section parameter
 $\sigma_{\rm SI}\sim 10^{-44} (n/5)^4 \,{\rm cm}^2 $
 (up to a QCD uncertainty of about one order of magnitude).
 As illustrated in fig.\fig{direct}, this is within the sensitivity of future experiments.

DM DM annihilations in the galactic halo into vector bosons can be resonantly enhanced,
giving indirect DM signals,
only for DM masses close to the values listed in eq.\eq{resonant}.
CC production of $\DM^\pm$ occurs at tree level with a much larger cross section,
that can exceed $10^{-34}\,{\rm cm}^2$,
but in ordinary situations it is forbidden kinematically.
We also discussed prospects for attempting CC direct DM detection
by accelerating an intense nuclear beam.

\appendix

\footnotesize

\paragraph{Acknowledgments}
We thank B. Bajc, R. Barbieri, V. Cavasinni, F. Cervelli, A. Delgado,
F. Gianotti, M. Caprio, G. Giudice, G. Isidori, A. Kievsky, D. McKinsey, A. Perrotta, V. Rychkov, M.Tamburini and P. Ullio for useful discussions and S. Pittel for communications.
The research of A.S.\ is supported in part by the European Programme
`The Quest For Unification', contract MRTN-CT-2004-503369 and by
the European Network of
     Theoretical Astroparticle Physics ILIAS/N6 under contract number
     RII3-CT-2004-506222.
The work of M.C. is supported in part by the USA DOE-HEP Grant
DE-FG02-92ER-40704. The work of N.F. is supported by a
joint Research Grant of the Italian Ministero dell'Istruzione,
dell'Universit\`a e della Ricerca (MIUR) and of the Universit\`a di
Torino within the {\sl Astroparticle Physics Project} and by a Research
Grant from INFN.

\centerline{\mini
This paper  was probably inspired by the recent anthropic trend, that
we prefer to acknowledge only in this Nicodemitic way.}

\begin{multicols}{2}
  
\end{multicols}


\begin{thebibliography}{nn}

  \bibitem{DM}
  For a recent review about
  supersymmetric DM see
  \art{G. Jungman, M. Kamionkowski, K. Griest}{Phys. Rep.}{267}{195}{1996}.



\bibitem{FT} The fine-tuning argument was invented to  motivate new physics at the weak scale.
    The other side of the coin was addressed, after LEP2 negative searches, in
  \art[hep-ph/9811386]{L. Giusti, A. Romanino, A. Strumia}{Nucl. Phys.}{B550}{3}{1999};
  \art[hep-ph/9905281]{R. Barbieri, A. Strumia}{Phys. Lett.}{B462}{144}{1999}.


\bibitem{strong} Strongly interacting, ``hadronized", Dark Matter is subject to a number of constraints analysed in \art{G.~D.~Starkman, A.~Gould, R.~Esmailzadeh and S.~Dimopoulos}{\PR}{D41}{3594}{1990} that leave only implausible windows open. 
The particular case of DM as a thermal relic  seems fully excluded by direct detection experiments and heavy isotopes searches.


\bibitem{decayDM}
R. Cen, Astrophys.J. 546 (2001)  L77;
M. Oguri, K. Takahashi, H. Ohno, K. Kotake, Astrophys.J. 597 (2003) 645;
K. Ichiki, P. M. Garnavich, T. Kajino, G. J. Mathews, M. Yahiro, Phys.Rev.
D68 (2003) 083518;
K. Takahashi, M. Oguri, K. Ichiki, Mon.Not.Roy.Astron.Soc. 352 (2004) 311

\bibitem{decayDM_CMB}
K. Ichiki, M. Oguri, K. Takahashi, Phys. Rev. Lett. 93 (2004) 071302.



\bibitem{Quasi-minimal}
  L.~J.~Hall, T.~Moroi and H.~Murayama,
  Phys.\ Lett.\ B {424} (1998) 305
  [hep-ph/9712515].
D.~Tucker-Smith and N.~Weiner,  Phys.\ Rev.\ D {72} (2005) 063509
  [hep-ph/0402065].




\bibitem{Antro}
\hepart[hep-th/0501082]{N.~Arkani-Hamed, S.~Dimopoulos, S.~Kachru}.
\hepart[hep-ph/0510064]{R. Mahbubani, L. Senatore}.
See also
\hepart[hep-ph/0510036]{G.M. Vereshkov et al.} and
\hepart[hep-ph/0510311]{M.~Masip, I.~Mastromatteo}.


\bibitem{AM}
 J.L.~Feng, T.~Moroi, L.~Randall, M.~Strassler and S.F.~Su,
  Phys.\ Rev.\ Lett.\   {83} (1999) 1731
  [hep-ph/9904250].
T.~Gherghetta, G.~F.~Giudice and J.~D.~Wells,
  Nucl.\ Phys.\ B {559} (1999) 27
  [hep-ph/9904378].



\bibitem{Tparity}
H.~C.~Cheng and I.~Low,
  JHEP {0408}, 061 (2004).


\bibitem{KK}
H.~C.~Cheng, J.~L.~Feng and K.~T.~Matchev,
  Phys.\ Rev.\ Lett.\  {89} (2002) 211301
  [hep-ph/0207125].


\bibitem{Coulomb}
C.A. Coulomb, ``Premi\`ere M\'emoire sur l'\'{E}le\-ctri\-citit\'e et
Magn\'etisme'', Histoire de l'Acad\'emie Royale des Sciences (1785) 569.



\bibitem{KT}
 E.~W.~Kolb and M.~S.~Turner, ``The Early Universe'', Addison-Wesley (1993).




\bibitem{altri}
 \art{M. Srednicki, R. Watkins, K.A. Olive}{\NP}{B310}{693}{1988}.
 \art{P. Gondolo, G. Gelmini}{\NP}{B360}{145}{1991}.

\bibitem{Burgess}
C.~P.~Burgess, M.~Pospelov and T.~ter Veldhuis,
  Nucl.\ Phys.\ B {619} (2001) 709 [hep-ph/0011335] and references therein.

\bibitem{WMAP}
\hepart[astro-ph/0603449]{D.N.~Spergel {\it et al.} (WMAP Science Team)}.





\bibitem{GoodWit}
  M.~W.~Goodman and E.~Witten,
  Phys.\ Rev.\ D {31} (1985) 3059.




\bibitem{CDMS}
CDMS collaboration, astro-ph/0509259.


\bibitem{future}
For the Xenon project see \art[astro-ph/0407575]{E. Aprile {\it et al.}}{Nucl.\ Phys.\ Proc.\ Suppl.}{138}{2005}{156}.
For the SuperCDMS project see \art[astro-ph/0503583]{CDMS-II collaboration}{
eConf}{C041213}{2529}{2004}.
Useful comparisons can be done using the tools in \url{dendera.berkeley.edu/plotter/entryform.html}.


\bibitem{Drees}
  \art{M. Drees, M. Nojiri}{\PR}{D48}{3483}{1993}.


\bibitem{coupl}
A. Bottino, F. Donato, N. Fornengo, S. Scopel,
Astropart. Phys. 18 (2002) 205; A. Bottino, F. Donato, N. Fornengo, S.
Scopel, Astropart. Phys. 13 (2000) 215.


\bibitem{HisanoDirect}
  J.~Hisano, S.~Matsumoto, M.~M.~Nojiri and O.~Saito,
  Phys.\ Rev.\ D {71} (2005) 015007  [hep-ph/0407168].


\bibitem{Pittel}
  See also \art{J. Engel, S. Pittel, P. Vogel}{Int. Journ. Mod. Phys.}{E1}{1}{1992}.

\bibitem{dama} R. Bernabei {\it et al.} (DAMA Collaboration), Riv. Nuovo Cim. {\bf 26 n. 1}, 1--73 (2003) 1-73 [astro-ph/0307403].


\bibitem{SUSYscan}
For scans in different SUSY models, see for instance:
\art[hep-ph/0505019]{S. Baek, D.G. Cerdeno, Y.G. Kim, P. Ko, C. Munoz},{JHEP}{0506}{017}{2005};
\art[hep-ph/0502001]{J. Ellis, K. Olive, Y. Santoso, V. Spanos}{Phys. Rev.}{D71}{095007}{2005};
\art[hep-ph/0412058]{A. Masiero, S. Profumo, P. Ullio}{Nucl. Phys.}{B712}{86}{2005};
\art[hep-ph/0303201]{U. Chattopadhyay, A. Corsetti, P. Nath}{Phys. Rev.}{D68}{035005}{2003}
\art[hep-ph/0302032]{J. Ellis, A. Ferstl, K.A. Olive, Y.D. Santoso}{Phys. Rev.}{D67}{123502}{2003};
\art[hep-ph/0305191]{H. Baer, C. Balazs, A. Belyaev, J. O'Farrill}{JCAP}{0309}{007}{2003}.
\art[hep-ph/0303130]{A. Lahanas, D. Nanopoulos}{Phys. Lett.}{B568}{55}{2003};
\art[hep-ph/0307303]{A. Bottino, F. Donato, N.Fornengo, S. Scopel}{Phys. Rev.}{D69}{037302}{2003};
\art[hep-ph/0208069]{Y. Kim, T. Nihei, L. Roszkowski, R. Ruiz de Austri}{JHEP}{12}{034}{2002};
\art[hep-ph/0010203]{A. Bottino, F. Donato, N.Fornengo, S. Scopel}{Phys. Rev.}{D63}{125003}{2001};
\art[hep-ph/9908427]{V.A. Bednyakov, H.V. Klapdor--Kleingrothaus}{Phys. Rev.}{D62}{043524}{2000};
\art[hep-ph/0001019]{E. Accomando, R. Arnowitt, B. Dutta, Y. Santoso}{Nucl. Phys.}{B585}{124}{2000};

\bibitem{HisanoAgain}
J.~Hisano, M.~Nagai, M.~M.~Nojiri and M.~Senami,
  AIP Conf.\ Proc.\  {\bf 805} (2006) 423 [hep-ph/0504068].

\bibitem{Hisano}
   J.~Hisano, S.~Matsumoto and M.~M.~Nojiri,
  Phys.\ Rev.\ D {67} (2003) 075014 [hep-ph/0212022].
J.~Hisano, S.~Matsumoto and M.~M.~Nojiri,
  Phys.\ Rev.\ Lett.\  {92} (2004) 031303  [hep-ph/0307216].
J.~Hisano, S.~Matsumoto, M.~M.~Nojiri and O.~Saito,
  Phys.\ Rev.\ D {71} (2005) 063528  [hep-ph/0412403].
J.~Hisano, S.~Matsumoto, O.~Saito and M.~Senami,
  hep-ph/0511118.




\bibitem{preHisano}
  L.~Bergstrom and P.~Ullio,
  Nucl.\ Phys.\ B {504} (1997) 27
  [hep-ph/9706232].
  Z. Bern, P. Gondolo and M. Perelstein, Phys. Lett. B 411 (1997) 86.




\bibitem{NR}
\art{W.E. Caswell, G.P. Lapage}{\PL}{167B}{437}{1986}.
\art{G.T. Bodwin, E. Braaten, G.P. Lapage}{\PR}{D51}{1125}{1995}.


\bibitem{amanda} J. Ahrens et al. (AMANDA Collaboration), Phys. Rev. D {66} (2002) 032006.

\bibitem{icecube} J. Lundberg, J. Edsjo, Phys.Rev. D {69} (2004) 123505.

\bibitem{DMnu}
\art[hep-ph/0205116]{P. R. Crotty}{\PR}{D66}{063504}{2002}.
\art[hep-ph/0506298]{M. Cirelli et al.}{Nucl. Phys.}{B727}{99}{2005}.


\bibitem{pbar1}  F. Donato, N.Fornengo, D. Maurin, P. Salati, R. Taillet,
Phys. Rev. D {69} (2003) 063501.

\bibitem{pamela} See: \url{wizard.roma2.infn.it/pamela}.

\bibitem{AMS} See: \url{ams.cern.ch/AMS}.

\bibitem{GAPS} K.~Mori, C.J.~Hailey, E.~A.~Baltz, W.W.~Craig, M.~Kamionkowski, W.~T.~Serber and P.~Ullio,
  Astrophys.\ J.\  {566} (2002) 604.

\bibitem{dbar0}  F. Donato, N. Fornengo, P. Salati, Phys.Rev. D {62} (2000) 043003.
\bibitem{profumo} H. Baer and S. Profumo, astro-ph/0510722.

\bibitem{GLAST} A. Morselli, Nucl. Phys. Proc. Suppl. {134} (2004) 127.

\bibitem{pieri} N. Fornengo, L. Pieri, S. Scopel, Phys. Rev. D {70} (2004) 103529.

\bibitem{VERITAS} T. C. Weekes et al., in Proc. of the 25th ICRC {5} (1997) 173.



\bibitem{STWY}
\art[hep-ph/0405040]{R.~Barbieri, A.~Pomarol, R.~Rattazzi and A.~Strumia}{\NP}{B703}{127}{2004}.
  G.~Marandella, C.~Schappacher and A.~Strumia,
  Nucl.\ Phys.\ B {715} (2005) 173
  [hep-ph/0502095].


\bibitem{LHC28}
For discussions about realistic LHC upgrades see
\art{G. Azuelos et al. ({\sc Atlas} collaboration)}{J. Phys.}{G28}{2453}{2002};
A. de Roeck,  talk at the
2003 International Workshop on Future Hadron Colliders,
web site \url{conferences.fnal.gov/hadroncollider}.
D. Denegri, talk at the 2005 Les Houches Workshop,
web site \url{lappweb.in2p3.fr/conferences/LesHouches}.


\bibitem{CHPT}
For a recent review of chiral perturbation theory see
  G.~Colangelo and G.~Isidori, hep-ph/0101264.


\end{thebibliography}
\end{document}